\documentclass[journal]{IEEEtran}

\usepackage{amsmath,amsfonts}
\usepackage{algorithmic}
\usepackage{algorithm}
\usepackage{array}
\usepackage[caption=false,font=normalsize,labelfont=sf,textfont=sf]{subfig}
\usepackage{textcomp}
\usepackage{stfloats}
\usepackage{url}
\usepackage{verbatim}
\usepackage{graphicx}
\usepackage{cite}
\usepackage{enumitem}
\usepackage{hyperref}
\usepackage{makecell}
\usepackage{etoolbox}
\robustify\bfseries
\usepackage{amssymb}
\usepackage{tabularx}
\usepackage{wasysym}
\usepackage{color,soul}
\usepackage{balance}
\usepackage{booktabs}
\usepackage{multirow}
\usepackage{siunitx}
\usepackage{arydshln}

\def\H{{\mathsf H}}
\def\T{{\mathsf T}}
\def\CC{{\mathbb C}}

\usepackage{caption}

\usepackage{amssymb}%
\usepackage{pifont}%

\usepackage{tikz}

\usepackage{flushend}

\begin{document}

\title{Mixture to Beamformed Mixture: Leveraging Beamformed Mixture as Weak-Supervision for Speech Enhancement and Noise-Robust ASR}

\author{Zhong-Qiu Wang and Ruizhe Pang
\thanks{Manuscript received June 13, 2025.
\textit{(Corresponding author: Zhong-Qiu Wang).}
}
\thanks{
Z.-Q. Wang and R. Pang are with the Department of Computer Science and Engineering, Southern University of Science and Technology, Shenzhen 518055, China (e-mail: \{wang.zhongqiu41,ruizhepang\}@gmail.com).}
}

\markboth{In submission}
{Shell \MakeLowercase{\textit{et al.}}: Bare Demo of IEEEtran.cls for IEEE Journals}
\maketitle

\begin{abstract}
In multi-channel speech enhancement and robust automatic speech recognition (ASR), beamforming can typically improve the signal-to-noise ratio (SNR) of the target speaker and produce reliable enhancement with little distortion to target speech.
With this observation, we propose to leverage beamformed mixture, which has a higher SNR of the target speaker than the input mixture, as a weak supervision to train deep neural networks (DNNs) to enhance the input mixture.
This way, we can train enhancement models using pairs of real-recorded mixture and its beamformed mixture, and potentially realize better generalization to real mixtures, compared with only training the models on simulated mixtures, which usually mismatch real mixtures.
Evaluation results on the real-recorded CHiME-4 dataset show the effectiveness of the proposed algorithm.
\end{abstract}


\IEEEpeerreviewmaketitle

\vspace{-0.2cm}
\section{Introduction}

\IEEEPARstart{D}{eep} learning has dramatically advanced speech enhancement \cite{WDLreview, Haeb-Umbach2019, Haeb-Umbach2020, Haeb-Umbach2025, Zheng2023, Araki2025, Xie2025}.
The most widely-adopted approach to date is supervised deep learning, where DNNs are trained on simulated pairs of clean speech and its mixture with interference signals (such as noises, reverberation, and concurrent speech) to predict the clean speech based on the simulated mixture \cite{WDLreview, Haeb-Umbach2019, Haeb-Umbach2020, Zheng2023, Araki2025, Xie2025, Haeb-Umbach2025}.
Although strong performance is observed on simulated mixtures, many studies report that, even with large-scale training \cite{WDLreview, Chen2016c, Zhang2023USES, Zhang2024USES2}, where massive mixtures are simulated to cover many variations during training, this approach has limited generalizability to real-recorded mixtures \cite{Cox2023ClaritySP2023, Vincent2016CHiME4Analysis, Watanabe2020CHiME6, Yu2022M2MeT, Wang2023MISP}.
This is largely because the current simulation techniques cannot simulate mixtures that are sufficiently realistic.
As a result, simulated training mixtures are often mismatched with real test mixtures, consequently resulting in limited generalizability.

One potential solution is leveraging real mixtures for model training.
However, real mixtures are typically unlabeled and labeling them at the sample level is nearly impossible.
Supervised learning based approaches hence cannot be applied.
In this context, unsupervised methods \cite{Wisdom2020MixIT,Tzinis2022REMIXT,Fujimura2021,Bando2021NeuralFCA,Aralikatti2022RAS,Wang2023UNSSOR,Wang2024USDnet} have been proposed.
However, due to not leveraging any supervision, their performance and application scenarios are often limited.

Various types of weak supervision such as the ones afforded by source prior models \cite{Zhang2018WeaklySupervisedASS, Xu2025ArrayDPS}, adversarially-trained discriminators \cite{Stoller2018}, target-sound classifiers \cite{Pishdadian2020FindStrength} and ASR models \cite{Chang2019MIMOSpeech} have been designed for training models directly on mixtures.
However, many types of weak supervision are not at the sample level and do not offer fine-grained supervision.
In multi-channel speech enhancement, beamforming can usually produce reliable enhancement to target speech and boost the SNR of the target speaker.
Intuitively, beamformed mixtures could be utilized as a \textit{sample-level} weak supervision to train DNNs to enhance the observed, lower-SNR input mixtures.

In this context, we propose \textit{mixture to beamformed mixture} (M2BM) training.
It builds upon the recent \textit{mixture to mixture} (M2M) algorithm \cite{Wang2024M2M}, where DNNs are trained to produce an estimate for target speech and another for non-target signals such that the two estimates, after being linearly filtered and summated, can approximate observed mixtures.
M2BM extends M2M by further encouraging that the two estimates can be linearly filtered and summated to approximate beamformed mixtures.
This way, we could leverage the weak-supervision in beamformed mixtures for enhancing the lower-SNR input mixtures.
Following SuperM2M \cite{Wang2024SuperM2M}, which combines M2M training on real mixtures with supervised learning on simulated mixtures, we combine M2BM training with supervised training, leading to SuperM2BM.
Evaluation results on the popular CHiME-$4$ dataset show the effectiveness of SuperM2BM.

\vspace{-0.2cm}
\section{Related Works}

There are studies \cite{Drude2019,Seetharaman2018,Tzinis2019,Zhang2022} leveraging conventional signal processing algorithms such as spatial clustering and beamforming to derive pseudo-labels for training separation or enhancement models directly on mixtures.
However, the DNN estimates are trained to directly approximate the pseudo-labels, which are typically not sufficiently accurate. Therefore, their performance is limited by that of conventional algorithms.
Differently, in M2BM, the two DNN estimates are trained to respectively approximate the target and non-target components in beamformed mixtures.

\begin{figure*}
  \centering
  \includegraphics[width=14cm]{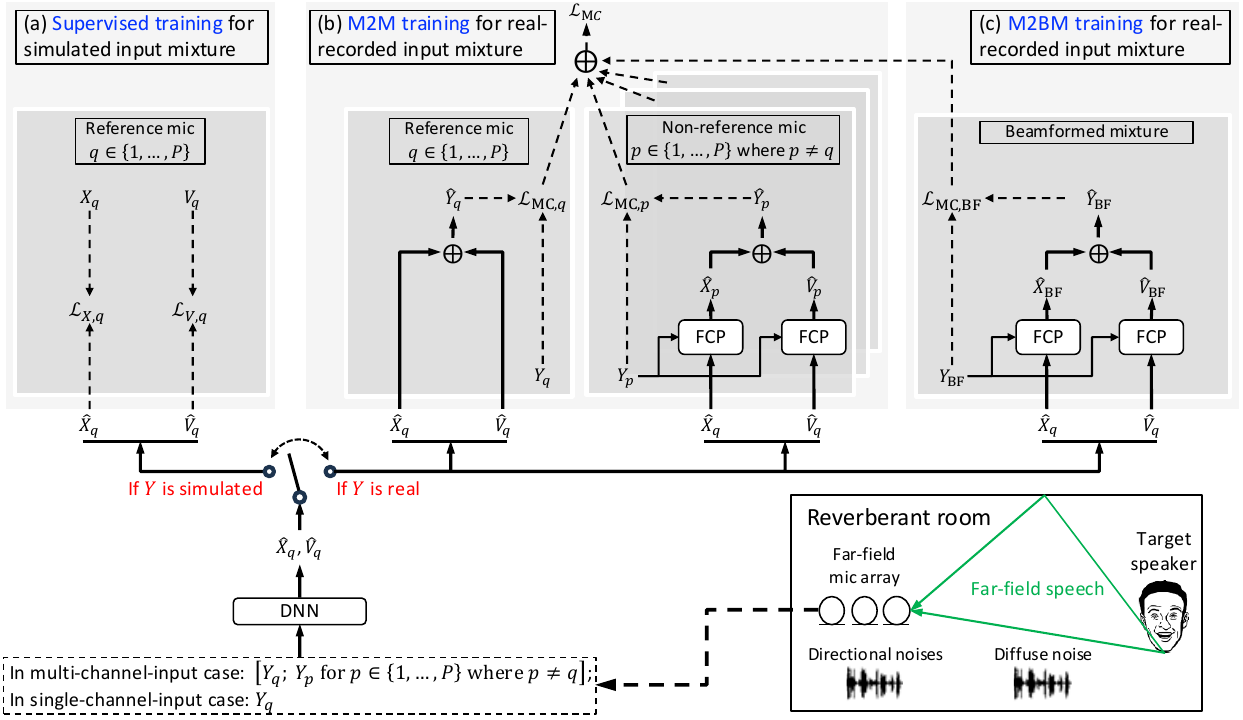}
  \vspace{-0.1cm}
  \caption{
  Illustration of SuperM2BM, which trains the same DNN-based enhancement model based on (a) supervised training if the input mixture is simulated; and (b) M2M training and (c) M2BM training if the input mixture is real-recorded.
  }\label{system_overview}
  \vspace{-0.5cm}
\end{figure*}

\vspace{-0.2cm}
\section{Physical Model and Objectives}

In a noisy-reverberant environment with a $P$-microphone far-field microphone array and a single target speaker (illustrated in the bottom-right plot of Fig. \ref{system_overview}), the physical model of the recorded far-field mixture can be formulated, in the short-time Fourier transform (STFT) domain, as follows:  
\begin{align}
Y_p(t,f) = X_p(t,f) + V_p(t,f), \label{phymodel_ff}
\end{align}
where $p\in \{1,\dots,P\}$ indexes the $P$ far-field microphones, $t$ indexes $T$ frames, $f$ indexes $F$ frequency bins, and $Y_p(t,f)$, $X_p(t,f)$, and $V_p(t,f)$ are respectively the mixture, reverberant speech of the target speaker, and reverberant signals of the other sources captured by microphone $p$ at time $t$ and frequency $f$.
When dropping indices $p$, $t$ and $f$, we refer to the corresponding spectrograms.
We designate microphone $q\in \{1,\dots,P\}$ as the reference microphone.
Assuming weak reverberation, we aim at suppressing environmental noises and reconstruct target reverberant speech at the reference microphone (i.e., $X_q$) based on the multi-channel input mixture $Y$.

\section{Review of SuperM2M}

SuperM2BM builds on SuperM2M \cite{Wang2024SuperM2M}, which combines supervised training on simulated mixtures and M2M training on real mixtures to train the same DNN for enhancement.
This section reviews SuperM2M.

\subsection{Review of Supervised Training}

Fig. \ref{system_overview}(a) illustrates supervised training on simulated mixtures.
The DNN is trained to predict the target and non-target signals at the reference microphone $q$.
The estimates are denoted as $\hat{X}_q$ and $\hat{V}_q$.
Since the mixture is simulated, we can define a supervised loss for each estimate:
\begin{align}
\mathcal{L}_{\text{X},q} &= \frac{1}{\sum\nolimits_{t,f} |Y_q(t,f)|} \sum\nolimits_{t,f} \mathcal{G}\Big( X_q(t,f), \hat{X}_q(t,f) \Big), \\ 
\mathcal{L}_{\text{V},q} &= \frac{1}{\sum\nolimits_{t,f} |Y_q(t,f)|}\sum\nolimits_{t,f} \mathcal{G}\Big( V_q(t,f), \hat{V}_q(t,f) \Big),
\end{align}
where, let $|\cdot|$ extracts magnitude and $\mathcal{R}(\cdot)$ and $\mathcal{I}(\cdot)$ respectively extract real and imaginary components, $\mathcal{G}(\cdot, \cdot)$ is defined as
\begin{align}
\mathcal{G}( a, b) = \big| \mathcal{R}(a) - \mathcal{R}(b) \big| +\big| \mathcal{I}(a) - \mathcal{I}(b) \big| +\big| |a| - |b| \big|. \label{L_RI_Mag}
\end{align}

\subsection{Review of M2M Training}

Fig. \ref{system_overview}(b) illustrates M2M training on real mixtures.
Since real mixtures are unlabeled,
we cannot apply supervised loss to penalize the DNN estimates $\hat{X}_q$ and $\hat{V}_q$.
In M2M training \cite{Wang2024M2M, Wang2024SuperM2M}, the two estimates are constrained such that they can be linearly filtered to reconstruct the mixture captured at each microphone, by optimizing the so-called mixture-constraint (MC) loss, which can be defined as
\begin{align}
\mathcal{L}_{\text{MC}} = \mathcal{L}_{\text{MC},q} + \frac{1}{P-1} \times \sum\nolimits_{p=1, p\neq q}^{P} \mathcal{L}_{\text{MC},p}, \label{MC_loss}
\end{align}
where $\mathcal{L}_{\text{MC},q}$ and $\mathcal{L}_{\text{MC},p}$ are respectively the MC losses at the reference microphone $q$ and non-reference microphone $p$.
$\mathcal{L}_{\text{MC},q}$ constrains that the two estimates should add up to the mixture at the reference microphone $q$ (i.e., $Y_q$):
\begin{align}
\mathcal{L}_{\text{MC},q} &= \sum\nolimits_{t,f} \mathcal{F} \Big( Y_q(t,f), \hat{Y}_q(t,f) \Big) \nonumber \\
&=\sum\nolimits_{t,f} \mathcal{F} \Big( Y_q(t,f), \hat{X}_q(t,f) + \hat{V}_q(t,f) \Big),
\label{MC_loss_ref}
\end{align}
where $\mathcal{F} \big( Y_q(t,f), \hat{Y}_q(t,f) \big ) = \frac{\mathcal{G}\big( Y_q(t,f), \hat{Y}_q(t,f) \big)}{\sum\nolimits_{t',f'} |Y_q(t',f')|}$, with $\mathcal{G}(\cdot, \cdot)$ defined in (\ref{L_RI_Mag}).
$\mathcal{L}_{\text{MC},p}$ constrains that the two estimates, after being linearly filtered, should add up to the mixture at each non-reference microphone $p$ (i.e., $Y_p$):
\begin{align}
&\mathcal{L}_{}{}_{\text{MC},p} = \sum\nolimits_{t,f} \mathcal{F}\Big( Y_p(t,f), \hat{Y}_p(t,f) \Big) \nonumber \\
&= \sum\nolimits_{t,f} \mathcal{F}\Big( Y_p(t,f), \hat{X}_p(t,f) + \hat{V}_p(t,f) \Big) \nonumber \\
&=\sum\nolimits_{t,f} \mathcal{F}\Big( Y_p(t,f), 
\hat{\mathbf{h}}_p(f)^\H \overline{\hat{\mathbf{X}}}_q(t,f) + \hat{\mathbf{r}}_p(f)^\H \overline{\hat{\mathbf{V}}}_q(t,f)
\Big). \label{MC_loss_nonref}
\end{align}
In (\ref{MC_loss_nonref}), $\hat{X}_p(t,f) = \hat{\mathbf{h}}_p(f)^\H \overline{\hat{\mathbf{X}}}_q(t,f)$, where $(\cdot)^{\H}$ computes Hermitian transpose, $\overline{\hat{\mathbf{X}}}_q(t,f)=\big[ \hat{X}_q(t-I+1,f),\dots,\hat{X}_q(t+J,f)\big]^\T \in \CC^{I+J}$ stacks a window of $I+J$ T-F units, and $\hat{\mathbf{h}}_p(f)\in \CC^{I+J}$ is the linear filter.
Similarly, $\hat{V}_p(t,f)=\hat{\mathbf{r}}_p(f)^\H \overline{\hat{\mathbf{V}}}_q(t,f)$, where $\overline{\hat{\mathbf{V}}}_q(t,f)=\big[ \hat{V}_q(t-I+1,f),\dots,\hat{V}_q(t+J,f)\big]^\T \in \CC^{I+J}$ and $\hat{\mathbf{r}}_p(f) \in \CC^{I+J}$.
The filter taps $I$ and $J$ are hyperparameters to tune.

To compute $\mathcal{L}_{\text{MC},p}$, the linear filters $\hat{\mathbf{h}}_p(f)$ and $\hat{\mathbf{r}}_p(f)$ need to be computed first.
Both of them can be computed by using the forward convolutive prediction (FCP) algorithm \cite{Wang2021FCPjournal}.
For example, $\hat{\mathbf{h}}_p(f)$ can be computed as follows:
\begin{align}\label{fcp_proj_mixture}
\hat{\mathbf{h}}_p(f)  =
\underset{\mathbf{h}_p(f)}{\text{argmin}}
\sum\limits_t \frac{\Big| Y_p(t,f) - \mathbf{h}_p(f)^\H \overline{\hat{\mathbf{X}}}_q(t,f) \Big|^2}{\hat{\lambda}_p(t,f)}.
\end{align}
$\hat{\lambda}$ is a weighting term defined as $\hat{\lambda}_p(t,f) = \xi\times \text{max}(|Y_p|^2) + |Y_p(t,f)|^2$, where $\xi$ (tuned to $10^{-2}$) floors the weighting term and $\text{max}(\cdot)$ extracts the maximum value of a power spectrogram.
The same method can be used to compute the other filter $\hat{\mathbf{r}}_p(f)$.
Notice that the optimization problem in (\ref{fcp_proj_mixture}) is quadratic, and a closed-form solution can be readily computed and plugged into (\ref{MC_loss_nonref}) to compute the loss and train the DNN.

\subsection{Review of SuperM2M Training}

SuperM2M \cite{Wang2024SuperM2M} combines supervised and M2M training, forming a semi-supervised approach that can train the same DNN-based enhancement model based on both simulated and real mixtures.
If the input mixture is simulated, supervised training is used, while if it is real, M2M training is used.
This way, the model can readily learn signal patterns from massive simulated mixtures, and, meanwhile, see real mixtures during training so that it can generalize better to real mixtures.

In SuperM2M \cite{Wang2024SuperM2M}, for real mixtures, the number of microphones used in input can be smaller than that used in loss computation.
For example, we can train a DNN using monaural input while computing the loss on all the microphones.
At run time, the DNN performs monaural enhancement. 

\section{SuperM2BM}

We propose SuperM2BM, which improves SuperM2M in Fig. \ref{system_overview}(a) and (b) by including a new loss defined on beamformed mixtures.
This section describes M2BM training, and how to obtain beamformed mixtures for M2BM training.

\subsection{Mixture to Beamformed Mixture (M2BM)}

Fig. \ref{system_overview}(c) illustrates M2BM training on real mixtures.
For each real training mixture, let us assume that we can first compute a beamforming result $Y_{\text{BF}}$.
We will describe how to compute it later in Section \ref{beamforming_description}.
Since, when the number of microphones is sufficiently large, $Y_{\text{BF}}$ would have a much higher SNR of the target speaker than the input mixture recorded by any of the microphones, it could be leveraged as a weak supervision to train the enhancement model.
In detail, we consider the beamforming result as a mixture signal recorded by a \textit{virtual microphone}, and compute an additional MC loss, which follows $\mathcal{L}_{\text{MC},p}$ in (\ref{MC_loss_nonref}), for model training:
\begin{align}
&\mathcal{L}_{}{}_{\text{MC,BF}} = \sum\nolimits_{t,f} \mathcal{F}\Big( Y_\text{BF}(t,f), \hat{Y}_\text{BF}(t,f) \Big) \nonumber \\
&= \sum\nolimits_{t,f} \mathcal{F}\Big( Y_\text{BF}(t,f), \hat{X}_\text{BF}(t,f) + \hat{V}_\text{BF}(t,f) \Big) \nonumber \\
&=\sum\nolimits_{t,f} \mathcal{F}\Big( Y_\text{BF}(t,f), \hat{\mathbf{h}}_\text{BF}(f)^\H \overline{\hat{\mathbf{X}}}_q(t,f) + \hat{\mathbf{r}}_\text{BF}(f)^\H \overline{\hat{\mathbf{V}}}_q(t,f)
\Big), \label{MC_loss_bf}
\end{align}
where $\overline{\hat{\mathbf{X}}}_q(t,f)$ and $\overline{\hat{\mathbf{V}}}_q(t,f)$ are defined in the same way as those in (\ref{MC_loss_nonref}), and the linear filters are computed in the same way as (\ref{fcp_proj_mixture}).
Building on the loss in (\ref{MC_loss}), the training loss for real mixtures becomes
\begin{align}
\mathcal{L}_{\text{MC}} = \mathcal{L}_{\text{MC},q} + \frac{1}{P-1} \times \sum\nolimits_{p=1, p\neq q}^{P} \mathcal{L}_{\text{MC},p} + \mathcal{L}_{\text{MC,BF}}.\label{MC_BF_loss}
\end{align}

\subsection{Deriving Beamformed Mixtures for M2BM Training}\label{beamforming_description}

So far, we assume that a beamforming result can be first computed from real-recorded mixtures and later used for M2BM training.
We employ the method proposed in \cite{Wang2020chime} to compute the beamformer.
We first train a monaural supervised speech enhancement model based on simulated mixtures to predict the target and non-target signals, following the method shown in Fig. \ref{system_overview}(a).
Next, we apply the model to enhance the real mixture captured by each microphone in the training set, obtaining $\hat{\mathbf{X}}(t,f)=[\hat{X}_1(t,f), \dots, \hat{X}_P(t,f)]^\T$ and $\hat{\mathbf{V}}(t,f)=[\hat{V}_1(t,f), \dots, \hat{V}_P(t,f)]^\T$.
Assuming that sound sources are non-moving, we compute time-invariant spatial covariance matrices for the target and non-target signals
via
\begin{align}
\hat{\boldsymbol{\Phi}}_{\text{X}}(f) &= \sum\nolimits_t \hat{\mathbf{X}}(t,f) \hat{\mathbf{X}}(t,f)^{\H}, \\
\hat{\boldsymbol{\Phi}}_{\text{V}}(f) &= \sum\nolimits_t \hat{\mathbf{V}}(t,f) \hat{\mathbf{V}}(t,f)^{\H},
\end{align}
where $(\cdot)^{\H}$ computes Hermitian transpose.
Next, we compute the relative transfer function $\hat{\mathbf{c}}_q(f)$ as follows:
\begin{align}
\mathbf{r}(f) &= \mathcal{P}\big(\hat{\boldsymbol{\Phi}}_{\text{X}}(f)\big), \\
\hat{\mathbf{c}}_q(f) &= \mathbf{r}(f) / r_q(f),
\end{align}
where $\mathcal{P}(\cdot)$ extracts the principal eigenvector.
A time-invariant minimum variance distortionless response (MVDR) beamformer \cite{Gannot2017,Haeb-Umbach2019,Haeb-Umbach2025} is then computed as
\begin{align}
\hat{\mathbf{w}}_q(f) &= \frac{\hat{\boldsymbol{\Phi}}_{\text{V}}(f)^{-1} \hat{\mathbf{c}}_q(f)}{ \hat{\mathbf{c}}_q(f)^\H \hat{\boldsymbol{\Phi}}_{\text{V}}(f)^{-1} \hat{\mathbf{c}}_q(f) }, \label{MVDR_equation}
\end{align}
and the beamforming result is obtained by
\begin{align}
Y_{\text{BF}}(t,f) = \hat{\mathbf{w}}_q(f)^{\H} \mathbf{Y}(t,f).
\end{align}
After obtaining beamformed mixtures for all the real training mixtures, we leverage them to train the SuperM2BM system, by including the $\mathcal{L}_{}{}_{\text{MC,BF}}$ loss defined in (\ref{MC_loss_bf}).

Notice that the beamformed mixtures are computed by using a DNN trained on simulated mixtures in a supervised way.
The trained DNN itself often exhibits limited generalizability to real mixtures, as the immediate outputs of the DNN (i.e., $\hat{\mathbf{X}}$ and $\hat{\mathbf{V}}$) are often not accurate on real mixtures \cite{Cox2023ClaritySP2023, Vincent2016CHiME4Analysis, Watanabe2020CHiME6, Yu2022M2MeT, Wang2023MISP}.
Nonetheless, many existing studies \cite{WDLreview,Yoshioka2015,Heymann2015,Zhang2017a,Haeb-Umbach2019,Haeb-Umbach2020,Haeb-Umbach2025} have shown that linear beamforming results derived based on the DNN estimates usually exhibit little distortion to target speech and can often result in much better ASR performance.
This is largely because the beamforming result is obtained via linear filtering, which can introduce little distortion to target speech.
Although time-invariant beamforming typically cannot suppress non-target signals aggressively, it can reliably improve the SNR of the target speaker.
We could hence utilize the beamformed mixtures for M2BM training.

Like SuperM2M, SuperM2BM, once trained, does not need to compute beamformed mixtures for inference.
It just runs feed-forwarding once and uses $\hat{X}_q$ as the enhancement result.

\section{Experimental Setup}

We evaluate SuperM2BM based on CHiME-$4$ \cite{Barker2015,Vincent2016CHiME4Analysis,Barker2017CHiME3}, the most popular dataset to date for robust ASR and speech enhancement.
It contains both simulated (SIMU) and real-recorded (REAL) mixtures for training and evaluation.
The real mixtures are recorded by using a tablet mounted with $6$ microphones (with the second one on the rear and the other $5$ facing front) in $4$ representative environments including buses, streets, pedestrian areas and cafeteria, where room reverberation and strong non-stationary, multi-source noises can naturally co-exist.
During data collection, the target speaker hand-holds the tablet and utters text prompts shown on the screen of the tablet.
In the training, validation and test sets, there are respectively $7,138$, $1,640$ and $1,320$ simulated mixtures, and $1,600$, $1,640$ and $1,320$ real mixtures.
Note that CHiME-$4$ contains weak reverberation, and the challenge is mainly in noise suppression.
The sampling rate is $16$ kHz.

\begin{figure*}
  \centering  
  \includegraphics[width=15cm]{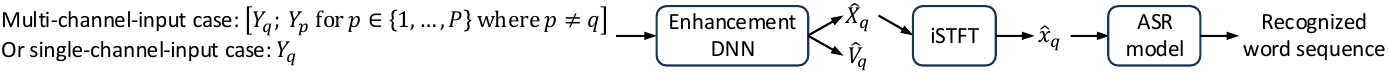}
  \vspace{-0.1cm}
  \caption{
  Robust ASR pipeline.
  }\label{enh_asr_pipeline_figure}
  \vspace{-0.6cm}
\end{figure*}

Fig. \ref{enh_asr_pipeline_figure} illustrates the evaluation pipeline for robust ASR, where enhanced speech is fed to a pre-trained backend ASR model for recognition.
We employ the strongest ASR model \cite{Chang2022E2EIntegration}
reported to date on CHiME-$4$. It is an encoder-decoder transformer model trained on WavLM features \cite{Chen2022WavLM}, using a transformer language model for decoding.

For enhancement, the STFT window size is $32$ ms, hop size $8$ ms, and the square root of the Hann window is used as the analysis window.
TF-GridNet \cite{Wang2023TFGridNet}, which has shown strong performance in recent speech separation benchmarks, is employed as the DNN architecture.
Using the symbols defined in Table I of \cite{Wang2023TFGridNet}, we set its hyperparameters to $D=128$, $B=4$, $I=1$, $J=1$, $H=200$, $L=4$ and $E=4$.
It is trained via complex spectral mapping \cite{Tan2020,Wang2020chime} to predict the real and imaginary (RI) components of the target and non-target signals based on the RI components of the input mixtures.
The filter taps $I$ and $J$ used to define $\overline{\hat{\mathbf{X}}}_q(t,f)$ and $\overline{\hat{\mathbf{V}}}_q(t,f)$ in (\ref{MC_loss_nonref}) and (\ref{MC_loss_bf}) are respectively tuned to $20$ and $1$.
For each real training mixture, we use all the $6$ microphones to compute the M2M loss, while the input to the DNN can be configured to $1$-, $2$- or $6$-channel.
The beamformed mixture is computed based on the front five microphones, considering that the rear microphone has a much lower SNR \cite{Barker2017CHiME3}.
Word error rates (WER) and DNSMOS OVRL \cite{K.A.Reddy2021} are used as the evaluation metrics.

\section{Evaluation Results}

Let us first look at the ASR results in Table \ref{results_CHiME4_official}.
Comparing row 2a, 4a and 6a with row $0$, we observe that purely-supervised learning based systems (i.e., training via Fig. \ref{system_overview}(a)) have severe generalization issues to the real test data (especially in the $2$- and $6$-channel input cases), although they performed reasonably well on the simulated test data.
Compared with purely-supervised systems, SuperM2M, trained by combining Fig. \ref{system_overview}(a) and (b), clearly improves the ASR performance on the real data, confirming the effectiveness of including real mixtures for model training.
By including beamformed mixtures for loss computation, SuperM2BM, which combines the training methods in Fig. \ref{system_overview}(a), (b) and (c), further improves the ASR performance of SuperM2M.
This suggests the effectiveness of M2BM.
In addition, SuperM2BM outperforms IRIS \cite{Chang2022E2EIntegration} and MultiIRIS \cite{Masuyama2023SE}, the previous best.

Row $7$ reports the ASR results of the beamformed test mixtures, which can indicate the quality of the beamformed mixtures used for M2BM training.
The monaural supervised model in row $2$a is used to derive the beamformed mixtures, using the method described in Section \ref{beamforming_description}.
Comparing $6$c and $7$, we observe that $6$-channel SuperM2BM obtains ASR results comparable to the beamforming method.
From $2$b and $2$c, and $4$b and $4$c, we can see that the ASR improvement brought by using the beamformed mixtures for training is clear.

Table \ref{results_CHiME4_official} also reports the DNSMOS scores.
From $2$b and $2$c, $4$b and $4$c, and $6$b and $6$c, we observe that SuperM2BM obtains comparable or better DNSMOS scores than SuperM2M.
This confirms the effectiveness of using beamformed mixtures for training.
Comparing $2$c, $4$c and $6$c with $7$, we observe that SuperM2BM obtains clearly better or comparable DNSMOS scores.
This suggests that the enhancement performance of SuperM2BM is not limited by that of beamformed mixtures.

\begin{table}[t]
\scriptsize
\centering
\captionsetup{justification=centering}
\sisetup{table-format=2.2,round-mode=places,round-precision=2,table-number-alignment = center,detect-weight=true,detect-inline-weight=math}
\caption{\textsc{Robust ASR and DNSMOS OVRL Results on CHiME-$4$.}}
\vspace{-0.1cm}
\label{results_CHiME4_official}
\setlength{\tabcolsep}{3.5pt}
\resizebox{1.0\columnwidth}{!}{
\begin{tabular}{
l %
l %
S[table-format=1,round-precision=0] %
S[table-format=1.2,round-precision=2] %
S[table-format=1.2,round-precision=2] %
S[table-format=1.2,round-precision=2] %
S[table-format=2.2,round-precision=2] %
S[table-format=1.2,round-precision=2] %
S[table-format=2.2,round-precision=2] %
}

\toprule

& & & \multicolumn{2}{c}{DNSMOS OVRL$\uparrow$} & \multicolumn{4}{c}{WER (\%)$\downarrow$} \\

\cmidrule(lr){4-5}\cmidrule(lr){6-9}

 & & {\multirow{2}{*}[-6pt]{\begin{tabular}[c]{@{}c@{}}Input\\\#mics\end{tabular}}} & {Val.} & {Test} & \multicolumn{2}{c}{Val.} & \multicolumn{2}{c}{Test} \\

\cmidrule(lr){4-4}\cmidrule(lr){5-5}\cmidrule(lr){6-7}\cmidrule(lr){8-9}

Row & System & & {REAL} & {REAL} & {SIMU} & {REAL} & {SIMU} & {REAL} \\

\midrule

0 & Mixture \cite{Chang2022E2EIntegration} & 1 & 1.5207437948951077 & 1.3926635820452578 & 5.93 & 4.03 & 8.25 & 4.47 \\

\midrule

1 & IRIS \cite{Chang2022E2EIntegration} & 1 & {-} & {-} & \bfseries 3.16 & 2.03 & \bfseries 6.12 & 3.92 \\
[0.5ex]\hdashline\noalign{\vskip 0.5ex}
2a & Supervised & 1 & \bfseries 3.332486674127914 & \bfseries 3.1798154060996557 & 3.525073746312685 & 2.1497842840812714 & 8.050056032872618 & 4.825073567191368 \\
2b & SuperM2M & 1 & 3.296253110599874 & 3.139178983417608 & 3.289085545722714 & 2.050223090821933 & 6.92005976839746 & 3.596618244663459 \\
2c & SuperM2BM & 1 & 3.245069436202774 & 3.1384176215215285 & 3.392330383480826 & \bfseries 1.8179136398834766 & 6.803324617108704 & \bfseries 3.101499369424074 \\

\midrule

3 & MultiIRIS \cite{Masuyama2023SE} & 2 & {-} & {-} & 2.04 & 1.66 & \bfseries 2.04 & 2.65 \\
[0.5ex]\hdashline\noalign{\vskip 0.5ex}
4a & Supervised & 2 & \bfseries 2.9906165635073636 & \bfseries 2.828943002485689 & 1.5449852507374632 & 11.92521848150743 & 2.2926783713111694 & 23.177168480545565 \\
4b & SuperM2M & 2 & 1.8077410329256147 & 1.6365336199802238 & 1.5671091445427728 & 2.7139643792175226 & 2.2226372805379157 & 2.8539399318043814 \\
4c & SuperM2BM & 2 & 2.7325734786410423 & 2.5059621948378825 & \bfseries 1.4601769911504425 & \bfseries 1.3975441572329363 & 2.175943220022413 & \bfseries 1.933766173104769 \\

\midrule

5 & MultiIRIS \cite{Masuyama2023SE} & 6 & {-} & {-} & 1.22 & 1.33 & \bfseries 1.24 & 1.77 \\
[0.5ex]\hdashline\noalign{\vskip 0.5ex}
6a & Supervised & 6 & 2.38247107638841 & 2.110373807859522 & \bfseries 0.8296460176991151 & 41.76038939488919 & 1.3121031004856181 & 67.25676117520669 \\
6b & SuperM2M & 6 & 1.8350843802276555 & 1.6250243654414271 & \bfseries 0.8333333333333333 & 2.4152807994395075 & 1.3354501307433694 & 2.2607314680741744 \\
6c & SuperM2BM & 6 & \bfseries 2.476433792507206 & \bfseries 2.1702370147640093 & 0.8517699115044249 & \bfseries 1.2500460931450274 & 1.3447889428464699 & \bfseries 1.5647624830678688 \\

\midrule

  & $1$ch supervised \\
7 & +$5$ch beamform. & 5 & 2.4840799628085866 & 2.189826124431661 & 0.9328908554572272 & 1.2205464803274457 & 1.386813597310422 & 1.5460787519267598 \\

\bottomrule
\multicolumn{9}{l}{\textit{Notes}: The SuperM2M results reported here differ from the ones in \cite{Wang2024SuperM2M}, since close-talk}\\
\multicolumn{9}{l}{mixtures are assumed available in \cite{Wang2024SuperM2M} and used to derive an additional loss for training,}\\
\multicolumn{9}{l}{while this study assumes that close-talk mixtures are not available for training.}
\end{tabular}
}
\vspace{-0.5cm}
\end{table}

\section{Conclusions}

We have proposed M2BM, which leverages beamformed mixture as weak supervision to train speech enhancement models on unlabeled real mixtures.
We have combined M2BM with supervised learning on simulated mixtures, yielding SuperM2BM.
Evaluation results on the CHiME-$4$ dataset show that SuperM2BM can effectively exploit the weak supervision in beamformed mixtures to improve generalizability.
In closing,  we highlight that, besides simulated mixtures, SuperM2BM only requires a set of multi-channel unlabeled real mixtures for training.
In practical application development, one could just record some real mixtures from the target domain and leverage SuperM2BM to realize better generalizability.

\bibliographystyle{IEEEtran}
\bibliography{references.bib}

\end{document}